# Graphene Ripples as a Realization of a Two-Dimensional Ising Model: A Scanning Tunneling Microscope Study

J.K. Schoelz,[1] P. Xu,[2] V. Meunier,[3] P. Kumar[1], M. Neek-Amal[4], P.M. Thibado,[1,*] and F.M. Peeters[4]

[1]*Department of Physics, University of Arkansas, Fayetteville, Arkansas 72701, USA*
[2]*Laboratory for Physical Sciences, University of Maryland, College Park, Maryland 20740, USA*
[3]*Department of Physics, Applied Physics, and Astronomy, Rensselaer Polytechnic Institute, Troy, New York 12180, USA*
[4]*Department of Physics, University of Antwerpen, Groenenborgerlaan 171, B-2020 Antwerpen, Belgium*

Ripples in pristine freestanding graphene naturally orient themselves in an array that is alternately curved-up and curved-down; maintaining an average height of zero. Using scanning tunneling microscopy (STM) to apply a local force, the graphene sheet will reversibly rise and fall in height until the height reaches 60-70% of its maximum at which point a sudden, permanent jump occurs. We successfully model the ripples as a spin-half Ising magnetic system, where the height of the graphene is the spin. The permanent jump in height, controlled by the tunneling current, is found to be equivalent to an antiferromagnetic-to-ferromagnetic phase transition. The thermal load underneath the STM tip alters the local tension and is identified as the responsible mechanism for the phase transition. Four universal critical exponents are measured from our STM data, and the model provides insight into the statistical role of graphene's unusual negative thermal expansion coefficient.

PACS numbers: 68.65.Pq, 82.37.Gk, 68.37.Ef, 05.50.+q

## I. INTRODUCTION

The long-range stability of any two-dimensional (2D) crystal is considered impossible, the result of many years of laboratory research backed by the well-established theoretical work of Peierls, Landau, and the Mermin-Wagner theorem [1]. Therefore, deviations from planarity are essential to the stability of isolated graphene [2,3]. In fact, when pristine suspended graphene is imaged via transmission electron microscopy [4] or scanning tunneling microscopy (STM) [5], its topography resembles a network of adjacent hemispherical surfaces with openings turning alternately either upward or downward. Yet, this natural intrinsic roughening is not the only allowable configuration; it is possible to rearrange the ripples to achieve lattice distortions of a desired shape, size, or periodicity [6-10].

The Ising system, which was initially introduced to describe simple spin systems, has proven to be useful in many areas of physics, from statistical mechanics to biophysics. Given the coupled two-state nature of graphene ripples and their potential for collective behavior, the celebrated and versatile Ising model might, in principle, be applicable. While Onsager's famous 2D solution was initially applied to the spontaneous magnetization of a 2D square lattice ferromagnet [11,12], through group renormalization [13] it has been established that a wide range of systems belong to the 2D spin-half Ising universality class [14-21]. Bonilla and Carpio treated graphene's individual carbon atoms with the 2D Ising model [22,23] and reproduced the existence of ripples [24,25]. They found when carbon atoms are placed in a double-well potential at each lattice site, the nonlinear force (plus added noise that effectively breaks meta-stability) produces stable ripples after a short transient period.

In this Letter, we show that pristine freestanding graphene undergoes transitions from a flexible state to a rigid state, consistent with a general solution of the 2D Ising universality class. This conclusion is reached from our experiments measuring freestanding graphene's perpendicular displacement when both a local electric field and local heating are applied with STM. Our study provides a new 2D Ising framework for understanding the role of graphene's ripples.

## II. EXPERIMENTAL METHODS

The graphene sample was grown using chemical vapor deposition, then transferred onto a 2000-mesh, ultrafine copper grid having a lattice of square holes 7.5 μm wide with bar supports 5 μm wide. An Omicron ultrahigh-vacuum (base pressure is $10^{-10}$ mbar) low-temperature model STM, operated at room temperature, was used with tips manufactured in-house [26]. STM images of freestanding graphene, as well as constant-current (feedback on) tip height versus bias voltage and setpoint current measurements were acquired. During a constant-current, tip-height measurement, a topography scan is made (typically only over an area of 0.1 nm by 0.1 nm), the imaging scanner is paused at one point, and the feedback loop remained operational. Assuming the sample is stationary, this process indirectly probes its density of states (DOS) [27,28]. A second interaction is also taking place, however, in which the tip bias induces an image charge in the grounded sample, resulting in an electrostatic attraction that increases with the bias and causing the sample to move towards the STM tip.

## III. EXPERIMENTAL RESULTS

To image freestanding graphene using STM, we first developed a method to convert it from a flexible state to a rigid state. To illustrate this method, five characteristic height-voltage, Z(V) measurements are displayed in Fig. 1. All measurements were acquired at the same sample location, in order of increasing current (as labeled). The three low-current curves (red) are characterized by a continuous, reversible increase in tip height (approximately 40 nm) as the tip bias is increased. Notice that the higher current, 0.5 nA data has a slightly larger displacement compared to the lower current, 0.1 nA data. This is due to the extra contraction occurring at higher currents as shown in our previous work [9,29]. The 1.0 nA curve (black) shows one small jump up around 0.5 V, then a sudden ~30 nm permanent jump follows at ~2.5 V, before reaching a plateau. Notice, the height change before the permanent jump is ~60% of the total height change. The high-current curve (blue, 5 nA) shows a total tip height change of only 3-4 nm over the entire tip bias range of 0.1 V to 3.22 V (i.e., the rigid state). This curve is displaced at the top of the 1.0 nA curve because the jump in height for the 1.0 nA trial was permanent (as indicated by the one-way direction arrows). Once in the rigid state, an STM image could be obtained from freestanding graphene as shown in Fig. 2(a) using a 4 nm black-to-white height scale. The characteristic honeycomb structure is visible throughout the image, and the overall topography features a wide ridge running diagonally from the bottom left corner to the top right corner. Note that these are difficult images to obtain because graphene is still very floppy by STM standards. To quantify the statistical properties of the STM image, a height-height correlation function, $<Z(x,y)Z(x+r_x,y+r_y)>$ was computed and is displayed in Fig. 2(b). This autocorrelation function is elongated in the same direction as the graphene ridge, and the correlation values are shown as a line profile in Fig. 2(c). The decay is modulated by small-scale oscillations caused by the presence of atomic corrugations. Half the correlation line profile is displayed on a log-log plot in Fig. 2(d). At about 1 nm, this curve drops sharply due to the edge of the STM image. A line having slope -0.25 is shown for comparison.

## IV. ISING MODEL

To model the experimental results, the thermodynamic behavior of the magnetization, $M$ of a modified 2D spin-half Ising magnet was simulated as a function of an external field having spatial dependence, $h(r)$ and temperature, $T$. The corresponding model Hamiltonian, $H$ can be written:

$$H = -J(M) \sum_{\langle ij \rangle} s_i s_j - h(r) \sum_i s_i \qquad (1)$$

where the Ising spins, $s_i$ represent ripples having either positive ($s_i > 0$) or negative ($s_i < 0$) curvature and $J(M)$ is the coupling energy between the nearest neighbor ripples whose value depends on the total magnetization,

$$M = \sum_i s_i \qquad (2)$$

and $h(r) = h_o e^{-r/\xi}$ represents the external field (in units of $J$) due to the STM tip, which we assume decays exponentially with length scale, $\xi$. We associate $J(M)$ to the elastic energy of the ripples. Local heating (due to the increasing tunneling current) contracts the graphene and increases $J(M)$ such that ripples are no longer favored. The last term in the Hamiltonian is related to the electrostatic energy between the tip and sample, which breaks the up-down symmetry.

In the floppy state of graphene, $J(M) = -1$ yields the desired antiferromagnetic coupling between nearest neighbor spins, while in the rigid state $J(M) = +2$ is used to provide the desired ferromagnetic coupling. We performed Monte Carlo simulations using the Metropolis algorithm at different temperatures (in units of $J/k_B$) on a hexagonal lattice with 96,000 sites but having an overall nearly square layout [30-32]. An individual spin, $s_i$ represents an entire ripple, having a diameter of about 10 nm, which contains ~1,000 carbon atoms, giving us a scale transformation compared to Bonilla and Carpio [23]. This scale change assumes that ripples exist and allows us to model the collective behavior of the ripples.

## V. NUMERICAL RESULTS

Five characteristic magnetization-field $M(h_o)$ simulation curves are displayed in Fig. 3(a). All simulations were carried out for strategic values of $J$, $T$, and $\xi$ (as labeled). The three lowest dashed curves (red) are characterized by an increase in magnetization with field, plus as the temperature is lowered the overall magnetization increases. Qualitatively, these are similar to the lower three experimental $Z(V)$ curves, with the height, bias voltage (note, we assume the electric field varies linearly with bias voltage) and tunneling current playing the role of the order parameter, external field, and temperature, respectively. However, notice that as the tunneling current increases, the temperature in the simulation decreases. The next simulation (black curve) has a small jump followed by a larger permanent jump (indicated with one-way arrows). During this simulation run the first jump occurred because we increased the field decay length $\xi$ from 28 to 35 (in lattice spacing units). The second jump occurs because the magnetization has reached 60-70% of its maximum, and at this point the overall nearest neighbor coupling $J(M)$ is changed into +2. This value of $J(M)$ and at this simulation temperature, the system is ferromagnetic below $T_c$. Therefore, even as the external field is lowered back to zero, the system stays ordered and follows the upper simulation dashed curve (blue) thereafter (as indicated by the arrows). The ordered simulations show excellent agreement with the high-current experimental $Z(V)$ curves. Real-space images extracted from the simulations at four different magnetizations are shown as insets in Fig. 3(a).

## VI. DISCUSSION

As we sweep through a wide range of voltages and then step through a large range of tunneling currents we are, in effect, hunting for the proper condition where freestanding graphene will change from a floppy trampoline-type geometry, as shown schematically in the upper model in Fig. 3(b), to a more rigid, larger, single-curvature type structure shown in the lower model of Fig. 3(b). The reason the system changes its configuration is highlighted in a simple cross-sectional illustration, but with more details, shown in Fig. 3(c). At low current and low voltage, the graphene model is in an antiferromagnetic state, but just above $T_c$, with an arrangement of alternately-oriented ripples, as shown with the dashed curve in the left image of Fig. 3(c). As the voltage is increased (solid curve), the ripples reverse their orientation and provide a mechanism for greater perpendicular displacement [33]. Next, as the current is increased, graphene is heated and contracts, as depicted in the shorter wavelength illustration shown with the dashed curve in the right image of Fig. 3(c). The contraction leads to a larger elastic energy build up (like a compressed spring), making the system more unstable to perpendicular movement. This is similar to Euler buckling when a system is under uniaxial compression [34] (i.e., a vertical column buckling under too much weight). When more voltage is applied at the higher tunneling currents, the system suddenly jumps to form a larger structure which is both rigid and stable. This final state is the ferromagnetic state below $T_c$ and is shown as the larger solid line in Fig. 3(c). Notice, that the role of temperature in our Ising model is two-fold. Temperature increases the entropy of the system as it would normally, however higher temperatures also cause a contraction which increases the internal energy of the system and ultimately drives the change in geometry.

A large number of additional Z(V,I) data sets were acquired from the freestanding graphene surface and across numerous samples, and results are in Fig. 4. All the current-voltage data pairs at which graphene transitions from flexible to rigid are shown as open circles. When the voltage range is reduced, we found the sudden permanent jump occurred at a much higher tunneling current setpoint; similarly when the voltage range was expanded, the jump then occurred at a much lower tunneling current setpoint. For the highest voltage range sweeps (0.01 to 10 V), graphene would sometimes tear, while for the lowest voltage range sweeps the jump to the rigid state would sometimes not occur. The overall result is consistent with a $1/I^2$ behavior, in which thermal contraction at higher currents permits a lower voltage to drive the system into the rigid state. Remarkably, this trend also mimics the behavior of the critical field as a function of temperature for some ferromagnetic systems (i.e., $h_c$ vs. T) [35]. The inset of Fig. 4, shows another, larger Z(V) data set showing the flexible state in red, a single jump Z(V) in black, and then several rigid Z(V) in blue. Notice, the jump occurs when the total height reaches 60-70% of the maximum.

The constant-current Z(V) data sets shown in Figs. 1 and 4, along with the STM image shown in Fig. 2, all provide strong evidence that this system can be described by the 2D Ising magnet Hamiltonian. However, quantifying a system's dimensionality and internal degrees of freedom requires the measurement of critical exponents [36]. Our large data set spanning current, voltage, and displacement made it possible to determine *four* of the six static 2D universal critical exponents, as listed in Table 1. Measuring just two of the critical exponents is enough to calculate the other six; so their interrelationships can be tested by measuring four. The pair correlation critical exponent $\eta$ is a measure of the average domain size at the critical point. It was measured from the decay of the autocorrelation function for several STM images similar to the one shown in Fig. 2(b), and our average value (see Table 1) is in good agreement with the 2D Ising prediction. The critical isotherm exponent $\delta$ characterizes the very slow, power-law increase in height with bias voltage along the critical isotherm (constant current). It is calculated using the Z(V) data curve just after the sudden permanent jump, and assumes that the electric field scales linearly with tip bias in the range of interest, as shown in Fig. 5(a). Next, the spontaneous polarization critical exponent $\beta$ describes the slow increase in height with temperature, in the ordered state and near the critical temperature. It is calculated using all the high-current data sets acquired just after the sudden permanent jump for all the tip biases in the region of the jump, while assuming (from resistive heating arguments) that the local temperature increases with $I^2$ for the graphene beneath the STM tip, as shown in Fig. 5(b) [37]. Please see our previous work for our method of modeling the sample heating [9,29]. Finally, the susceptibility critical exponent $\gamma$ is a measure of how the susceptibility, $\partial Z/\partial V$ changes with temperature near the critical point. It is calculated for all the high-current data sets acquired just after the sudden permanent jump, as shown in Fig. 5(c). This large set of critical exponents, associated with the second-order phase transition, are all within the 2D Ising universality class [38], and therefore provides a rigorous testament of freestanding graphene's 2D Ising behavior.

One fascinating aspect for each of these data sets is that the transition from the flexible state (above $T_c$) to the rigid state (below $T_c$) occurs with increasing current, that is, when heating up the sample, which is opposite to the usual 2D Ising magnet behavior. This is a consequence of graphene's unusual negative thermal expansion property (see W. Bao et al., in which the coefficient was measured above room temperature in a similar suspended configuration) [39-43]. When graphene is heated, the internal tension increases, and this changes the coupling between the nearest-neighbor ripples. In effect it alters the lowest energy configuration for graphene from the antiferromagnetic state to the bulged out ferromagnetic state.

## VII. SUMMARY

In summary, this study successfully applied the 2D magnetic Ising model to the technologically-important freestanding graphene system. *Four* universal 2D critical exponents were measured from STM data. Unexpectedly, a transition was observed between a flexible state and a rigid state as the sample was heated (which is opposite the 2D Ising system); this is explained in terms of the negative thermal expansion properties of graphene. We presented a model in which individual graphene ripples are the spins of a 2D magnet, with the distinction that the elastic energy of the ripples increases during heating due to thermal contraction.

## ACKNOWLEDGMENTS

This work was supported in part by ONR under grant N00014-10-1-0181 and NSF under grant DMR-0855358. F.M. Peeters and M. Neek-Amal were supported by the Flemish Science Foundation (FWO-Vl) and the Methusalem Foundation of the Flemish Government.

## REFERENCES

[1] N. D. Mermin and H. Wagner, Phys. Rev. Lett. **17**, 1133 (1966).
[2] A. Fasolino, J. H. Los, and M. I. Katsnelson, Nature Mater. **6**, 858 (2007).
[3] P. Le Doussal and L. Radzihovsky, Phys. Rev. Lett. **69**, 1209 (1992).
[4] J. C. Meyer, A. K. Geim, M. I. Katsnelson, K. S. Novoselov, D. Obergfell, S. Roth, C. Girit, and A. Zettl, Solid State Commun. **143**, 101 (2007).
[5] R. Zan, C. Muryn, U. Bangert, P. Mattocks, P. Wincott, D. Vaughan, X. Li, L. Colombo, R.S. Ruoff, B. Hamilton, and K.S. Novoselov, Nanoscale **4**, 3065 (2012).
[6] N. Levy, S. A. Burke, K. L. Meaker, M. Panlasigui, A. Zettl, F. Guinea, A. H. Castro Neto, and M. F. Crommie, Science **329**, 544 (2010).
[7] N. N. Klimov, S. Jung, S. Zhu, T. Li, C. A. Wright, S. D. Solares, D. B. Newell, N. B. Zhitenev, and J. A. Stroscio, Science **336**, 1557 (2012).
[8] T. Mashoff, M. Pratzer, V. Geringer, T. J. Echtermeyer, M. C. Lemme, M. Liebmann, and M. Morgenstern, Nano Lett. **10**, 461 (2010).
[9] P. Xu, Y. Yang, S.D. Barber, M.L. Ackerman, J.K. Schoelz, D. Qi, I.A. Kornev, L. Dong, L. Bellaiche, S. Barraza-Lopez, and P.M. Thibado, Phys. Rev. B **85**, 121406(R) (2012).
[10] P. Xu, M. Neek-Amal, S. D. Barber, J. K. Schoelz, M. L. Ackerman, P. M. Thibado, A. Sadeghi, and F. M. Peeters, Nature Comm. **5**, 3720 (2014).
[11] H. A. Kramers and G. H. Wannier, Phys. Rev. **60**, 252 (1941).
[12] L. Onsager, Phys. Rev. **65**, 117 (1944).
[13] K. G. Wilson, Rev. Mod. Phys. **47**, 773 (1975).
[14] T. D. Lee and C. N. Yang, Phys. Rev. **87**, 410 (1952).
[15] Z. Friedman, Phys. Rev. Lett. **36**, 1326 (1976).
[16] R. Hocken and M. R. Moldover, Phys. Rev. Lett. **37**, 29 (1976).
[17] S. Janssen, D. Schwahn, and T. Springer, Phys. Rev. Lett. **68**, 3180 (1992).
[18] L. P. Kadanoff, W. Gotze, D. Hamblen, R. Hecht, E.A.S. Lewis, V.V. Palciauskas, M. Rayl, J. Swift, D. Aspnes, and J. Kane, Rev. Mod. Phys. **39**, 395 (1967).
[19] V. P. LaBella, D. W. Bullock, M. Anser, Z. Ding, C. Emery, L. Bellaiche, and P. M. Thibado, Phys. Rev. Lett. **84**, 4152 (2000).
[20] Z. Ding, D. W. Bullock, P. M. Thibado, V. P. LaBella, and K. Mullen, Phys. Rev. Lett. **90**, 216109 (2003).
[21] Z. Ding, D. W. Bullock, P. M. Thibado, V. P. LaBella, and K. Mullen, Surf. Sci. **540**, 491 (2003).
[22] L. L. Bonilla and A. Carpio, J. Stat. Mech., P09015 (2012).
[23] L. L. Bonilla, A. Carpio, A. Prados, and R. R. Rosales, Phys. Rev. E **85**, 031125 (2012).
[24] L. L. Bonilla and A. Carpio, Phys. Rev. B **86**, 195402 (2012).
[25] A. O'Hare, F. V. Kusmartsev, and K. I. Kugel, Nano Lett. **12**, 1045 (2012).
[26] J. K. Schoelz, P. Xu, S. D. Barber, D. Qi, M. L. Ackerman, G. Basnet, C. T. Cook, and P. M. Thibado, J. Vac. Sci. Technol. B **30**, 033201 (2012).
[27] E. N. Voloshina, E. Fertitta, A. Garhofer, F. Mittendorfer, M. Fonin, A. Thissen, and Y. S. Dedkov, Scientific Reports **3**, 1072 (2013).
[28] B. Borca, S. Barja, M. Garnica, M. Minniti, A. Politano, J.M. Rodriguez-Garcia, J.J. Hinarejos, D. Farias, A.L. Vazquez de Parga, and R. Miranda, New Journal of Physics **12**, 093018 (2010).
[29] M. Neek-Amal, P. Xu, J. K. Schoelz, M. L. Ackerman, S. D. Barber, P. M. Thibado, A. Sadeghi, and F. M. Peeters, Nature Comm. **5**, 4962 (2014).
[30] N. Metropolis, A. W. Rosenbluth, M. N. Rosenbluth, A. H. Teller, and E. Teller, Journal of Chemical Physics **21**, 1087 (1953).
[31] W. K. Hastings, Biometrika **57**, 97 (1970).
[32] V. P. LaBella, D. W. Bullock, Z. Ding, C. Emery, W. G. Harter, and P. M. Thibado, J. Vac. Sci. Technol. A **18**, 1526 (2000).
[33] A. A. Taskin, A. N. Lavrov, and Y. Ando, Phys. Rev. Lett. **90**, 227201, 227201 (2003).
[34] J. Jin Wu, Nanotechnology **25**, 355402 (2014).
[35] C. Antoniak, M.E. Gruner, M. Spasova, A.V. Trunova, F.M. Romer, A. Warland, B. Krumme, K. Fauth, S. Sun, P. Entel, M. Farle, and H. Wende, Nature Communications **2**, 528 (2011).
[36] M. E. Fisher, Rep. Prog. Phys. **30**, 615 (1967).

[37] R. Wiesendanger, *Scanning Probe Microscopy and Spectroscopy: Methods and Applications* (Cambridge University Press, Cambridge, UK, 1994).
[38] R. K. Pathria and P. D. Beale, *Statistical Mechanics* (Elsevier, Oxford, UK, 2011), 3 edn.
[39] W. Bao, F. Miao, Z. Chen, H. Zhang, W. Jang, C. Dames, and C. N. Lau, Nature Nanotech. **4**, 562 (2009).
[40] N. Mounet and N. Marzari, Phys. Rev. B **71**, 205214 (2005).
[41] G. Savarino and M. R. Fisch, American Journal of Physics **59**, 141 (1991).
[42] M. Pozzo, D. Alfè, P. Lacovig, P. Hofmann, S. Lizzit, and A. Baraldi, Phys. Rev. Lett. **106**, 135501 (2011).
[43] K. V. Zakharchenko, M. I. Katsnelson, and A. Fasolino, Phys. Rev. Lett. **102**, 046808 (2009).

FIG. 1. Constant-current, tip-height vs. bias voltage Z(V) data sets on suspended graphene acquired using the labeled setpoint currents. The red curves indicate the flexible state which are reversible, the black curve is when the permanent jump occurred, and the blue curve indicates the rigid state. Curves are slightly offset from each other for clarity.

FIG. 2. (a) Constant-current, filled-state STM image over a 6 nm × 6 nm surface of freestanding graphene taken with V = 0.1 V and I = 1.0 nA. (b) Site-site correlation of (a). (c) Line profile taken from (b). (d) Half of the line in (c) on a log-log plot (circles) with solid line power law function ($r^{-\eta}$ with $\eta = 0.25$).

FIG.3. (a) Dashed lines show the isothermal magnetization of a 2D Ising magnet vs. field using a particular $J$, $T$, and $\xi$ as labeled. Curves are slightly offset from each other for clarity. Insets: Real-space simulation images at various magnetizations. (b) Rendered 3D molecular models of freestanding graphene illustrating the antiferromagnetic state (upper image) and local ferromagnetic state (lower image). (c) Schematic showing graphene ripples in cross-section at a low (left) and high (right) tunneling current for both low (dashed) and high (solid) bias voltages. Curved-up (-down) graphene ripples are spin-up (-down) elements.

FIG. 4. Large set of experimental ($I_c$, $V_c$) points for which graphene transitions from floppy to rigid. The solid line is a trend curve to aid the eye. Inset: Complete data set showing several flexible Z(V) curves (red), a single jump curve in black, and several rigid curves in blue. Curves are slightly offset from each other for clarity. The three critical exponents $\delta$, $\gamma$, and $\beta$ are calculated using the highlighted data in the region marked in the inset figure and using the formulas shown in Table 1.

FIG. 5. Plots used to determine the value of the critical exponents $\delta$, $\beta$, and $\gamma$. (a) $\text{Log}|Z-Z_c|$ vs $\text{Log}|V-V_c|$ shown as solid circles. The red line, shown for comparison, has the ideal slope of $1/\delta$ or $1/15$. (b) $\text{Log}|Z-Z_c|$ vs $\text{Log}|I^2-(I_c)^2|$ shown as solid circles. The red line, shown for comparison, has the ideal slope of $\beta$ or $1/8$. (c) $\text{Log}|\partial Z/\partial V|$ vs $\text{Log}|I^2-(I_c)^2|$ shown as solid circles. The red line, shown for comparison, has the ideal slope of $-\gamma$ or $-7/4$.

TABLE 1. Measured critical exponents with predicted 2D and 3D values. $V_c$ and $I_c$ are the voltage and current at which the jump of height $Z_c$ occurred for a particular data set.

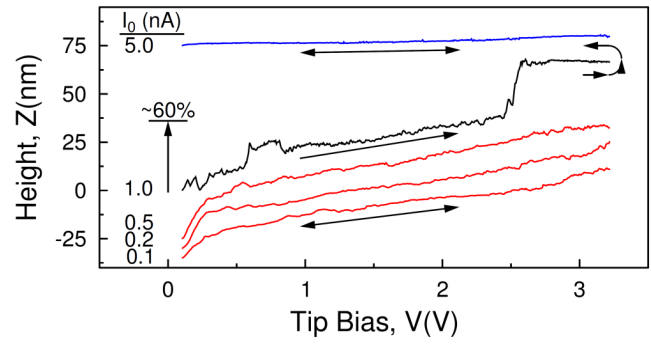

Figure 1

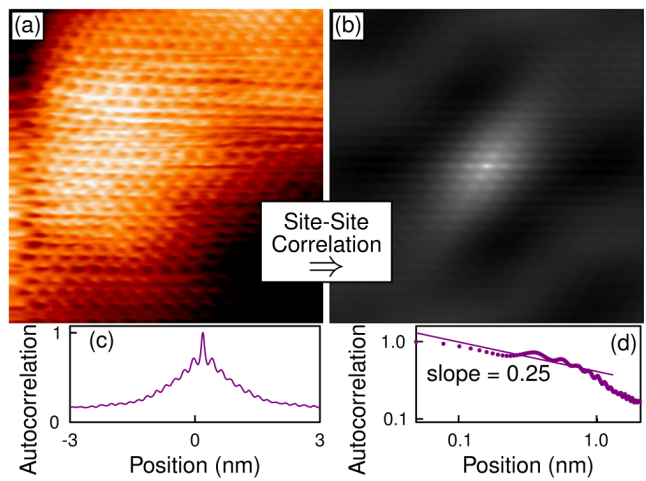

Figure 2

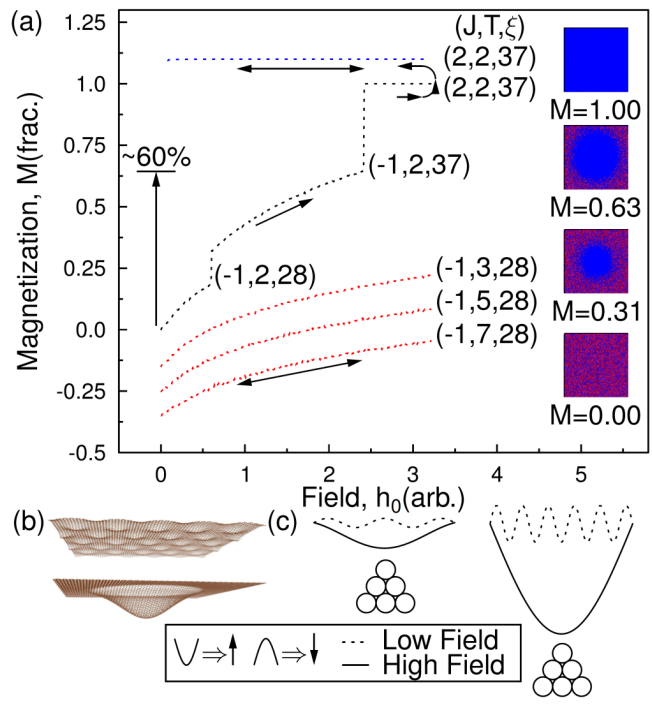

Figure 3

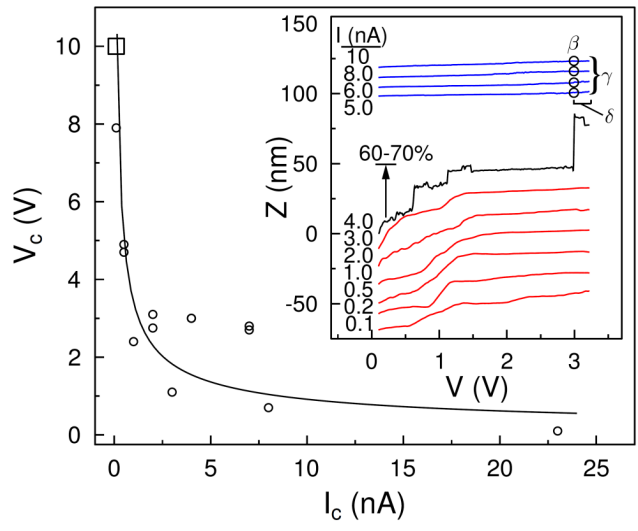

Figure 4

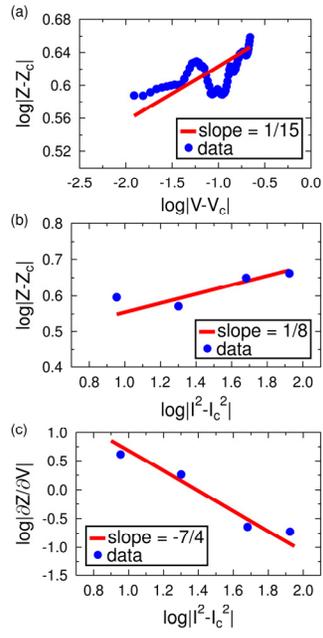

Figure 5

| Quantity | Ising Ferromagnet | Graphene Ripples | Measured Value | 2D | 3D |
|---|---|---|---|---|---|
| Pair Correlation Function ($T=T_c$) | $G(r) \sim r^{-\eta}$ | $G(r) \sim r^{-\eta}$ | $\eta = 0.27 \pm 0.03$ | 0.25 | 0.04 |
| Critical Isotherm ($T=T_c$) | $M \sim H^{1/\delta}$ | $|Z - Z_c| \sim |V - V_c|^{1/\delta}$ | $\delta = 15 \pm 3$ | 15 | 5 |
| Spontaneous Polarization | $M \sim |T - T_c|^\beta$ | $|Z - Z_c| \sim |I^2 - I_c^2|^\beta$ | $\beta = 0.12 \pm 0.03$ | 0.125 | 0.327 |
| Susceptibility ($\chi$) | $\partial M/\partial H \sim |T - T_c|^{-\gamma}$ | $|\partial Z/\partial V| \sim |I^2 - I_c^2|^{-\gamma}$ | $\gamma = 2.1 \pm 0.4$ | 1.75 | 1.23 |

Table 1